\documentclass[12pt]{article}
\usepackage{amssymb, amsthm, amsmath}
\usepackage{bm}
\usepackage{graphicx,psfrag,epsf}
\usepackage{enumerate}
\usepackage{natbib}
\usepackage{url} 
\usepackage{subfig}
\usepackage{color}
\usepackage{hyperref}
\usepackage{booktabs}

\newcommand{\blind}{1}

\begin{document}

	\def\spacingset#1{\renewcommand{\baselinestretch}%
		{#1}\small\normalsize} \spacingset{1}

	
	\if1\blind
	{
		\title{
			\bf Bayesian Modeling of \\Spatial Molecular Profiling Data\\ via Gaussian Process
		}
		\author{
			Qiwei Li\thanks{To whom correspondence should be addressed}, \thanks{These authors contributed equally to this work.}\\ 
			Department of Mathematical Sciences\\ The University of Texas at Dallas, Richardson Texas\\
			and \\
			Minzhe Zhang$^\dagger$, Yang Xie, Guanghua Xiao$^*$\\
			Quantitative Biology Research Center\\ Department of Population and Data Sciences\\ The University of Texas Southwestern Medical Center, Dallas, Texas
		}
		\maketitle
	} \fi

	\newpage
	\bigskip
	\begin{abstract}
		The location, timing, and abundance of gene expression (both mRNA and proteins) within a tissue define the molecular mechanisms of cell functions.  Recent technology breakthroughs in spatial molecular profiling, including imaging-based technologies and sequencing-based technologies, have enabled the comprehensive molecular characterization of single cells while preserving their spatial and morphological contexts. This new bioinformatics scenario calls for effective and robust computational methods to identify genes with spatial patterns. We represent a novel Bayesian hierarchical model to analyze spatial transcriptomics data, with several unique characteristics. It models the zero-inflated and over-dispersed counts by deploying a zero-inflated negative binomial model that greatly increases model stability and robustness. Besides, the Bayesian inference framework allows us to borrow strength in parameter estimation in a \textit{de novo} fashion. As a result, the proposed model shows competitive performances in accuracy and robustness over existing methods in both simulation studies and two real data applications. The related \texttt{R}/\texttt{C++} source code are available at https://github.com/Minzhe/BOOST-GP.
	\end{abstract}
	
	\noindent%
	\vfill
	
	\newpage
	\spacingset{1.45} 
	\section{Introduction}
	
	Investigating the spatial organization of cells, together with their mRNA and protein abundances, is essential to delineating how cells from different origins form tissues with distinctive structures and functions. Traditional molecular profiling technologies require tissue-dissociation, which leads to the loss of the spatial context of gene expression, while traditional imaging technologies can only measure the levels of several markers. Therefore, the morphological/spatial information of tissues and the high-throughput molecular profile, for a long time, were analyzed individually with little crosstalk. 
	
	Recent technology breakthroughs in spatial molecular profiling (SMP) \citep{zhang2020spatial}, including imaging-based methods such as seqFISH \citep{lubeck2014single,eng2019transcriptome} and MERFISH \citep{chen2015spatially} and sequencing-based methods such as spatial transcriptomics \citep{staahl2016visualization} and Slide-seq \citep{rodriques2019slide}, empower the comprehensive molecular characterization of cells while keeping their spatial and morphological contexts. These SMP technologies combine the spatial information of tissues and the high-throughput molecular profile, enabling mapping and measuring the gene expression or protein abundance of thousands of cells over a tissue slide simultaneously. It provides a comprehensive molecular characterization of single cells. 
	
	Many new questions can be studied with this powerful new technology available. One of the most immediate ones is to identify genes whose expressions display spatially correlated patterns, which we refer to as spatially variable (SV) genes. Such genes may reflect the tissue heterogeneity and underlying tissue structure that drive the differentiated expression across different spatial locations. Therefore they are potentially significant and may lead to new biological insights. Statistically, identifying SV genes is a new challenge to test the association between gene expression levels and their spatial coordinates with a null hypothesis of spatial invariance. Several methodologies have been recently proposed for this task. SpatialDE \citep{svensson2018spatialde} and SPARK \citep{sun2020statistical} are two that adopt a Gaussian process (GP), which serves as a natural fit for this problem because of its ability to model temporal \citep{roberts2013gaussian} or spatial \citep{diggle1998model} dependence via a pre-specified kernel. They share the same geostatistical modeling framework, with SPARK being more explicit in modeling count data, sample normalization, and $p$-value calibration.  Trendsceek \citep{edsgard2018identification} is another method that models the spatial gene expression of cells as a realization marked point process. Its core concept is to calculate some summary statistics for all pairwise points with respect to their distances and evaluate the significance under null distribution generated with random permutations. 
	
	All of these methods are for identifying SV genes in SMP data, while still leaving several issues unsolved. First, SpatialDE and Trendsceek transformed count data, which do not truly reflect the underlying data generative mechanism. Although SPARK used a Poisson distribution for modeling counts, the simple mean-variance relationship may not be sufficient in accounting for the over-dispersion observed in real sequencing data. Second, typical SMP data contains a large proportion of zero counts (e.g. $60\%$ – $99\%$), and none of the mentioned methods properly addressed the zero count, which may largely reduce the statistical power. Table \ref{Table 0} shows a brief summary of four widely analyzed SMP cohorts. Last but not least, SpatialDE and SPARK both employ GP to estimate spatial covariance but only at certain predefined length scales of the spatial kernels, which means the estimation is only an approximated solution. 
	\begin{table*}[!t]
		\caption{A summary of four widely analyzed spatial molecular profiling (SMP) cohort data\label{Table 0}} {
			\footnotesize
			\begin{tabular}{@{}lllccc@{}}
				\toprule 
				Name & Technique & \# of datasets & \# of locations $n$ & \# of genes $p$ & zero prop.\\\midrule
				Mouse olfactory bulb & STS & $12$ replicates  & $231$ -- $282$ spots  & $15,284$ -- $16,675$ & $89\%$ -- $90\%$ \\
				(MOB) \cite{staahl2016visualization}  &&&&&\\\hline
				Human breast cancer & STS & $4$ layers   & $251$ -- $264$ spots  & $14,789$ -- $14,929$ & $60\%$ -- $79\%$ \\
				 (BC) \cite{staahl2016visualization}  &&&&&\\\hline
				Mouse hypothalamus 
				& MERFISH & $31$ individuals   & $4,877$ -- $6,000$ cells  & $160$ -- $161$ & $59\%$ -- $68\%$ \\
				\cite{moffitt2018molecular}  &&&&&\\\hline
				Mouse hippocampus & seqFISH & $21$ fields & $97$ -- $362$ cells & $249$ & $1\%$ -- $18\%$  \\
				\cite{shah2016situ}  &&&&&\\\bottomrule
			\end{tabular}
		}{Abbreviations: STS is spatial transcriptomics sequencing; MERFISH is multiplexed error-robust fluorescence \textit{in situ} hybridization; seqFISH is sequential fluorescence \textit{in situ} hybridization.}
	\end{table*} 
	
	To address the issues mentioned above, we developed Bayesian mOdeling Of Spatial Transcriptomics data via Gaussian Process (BOOST-GP), which integrates the GP to capture the spatial correlation. Note that although the name only includes spatial transcriptomics data, the proposed method can be applied to analyze other SMP data without modification. BOOST-GP directly models count data using a negative binomial distribution, which was also adopted by widely used RNA-seq analysis tools such as DESeq2 \citep{love2014moderated} and edgeR \citep{robinson2010edger} to account for the over-dispersion observed in real sequencing data. Previous studies in single-cell RNA-seq (scRNA-seq) data analysis have shown that accounting for the large proportion of zero counts in the model significantly improves the model fitting and accuracy in identifying differentially expressed genes \citep{kharchenko2014bayesian,finak2015mast,lun2016pooling}. One novelty of this study is to explicitly accounts for the excess zeros through a zero-inflated negative binomial model. Furthermore, it uses a Bayesian inference framework to improve parameter estimations and quantify uncertainties. We demonstrated the advantages of BOOST-GP in robust inference across various spatial patterns and minimally affected performance when data contains zero inflation in the simulation study. We also applied BOOST-GP to two real spatial transcriptomics datasets, and BOOST-GP shows an outstanding sensitivity-specificity balance compared to existing methods.
	
	The rest of this article is arranged as follows. In Section \ref{model}, we formulate the probabilistic generative model, explore the parameters structure and deduce the posterior distribution. In Section \ref{model_fitting}, we present the schematic procedure and demonstrate the Markov chain Monte Carlo algorithms. In Section \ref{simulation} and \ref{real_data}, we evaluate BOOST-GP via both simulated and real data. The last section concludes the article and proposes some potential future work.

\section{Model}\label{model}
Let a $n$-by-$p$ count matrix $\bm{Y}$ denote the molecular profile generated by a spatially resolved transcriptomics technology based on either a sequencing or imaging-based method. The former can measure tens of thousands of genes on spatial locations consisting of a couple of hundred single cells (i.e. $p\gg n$), while the latter directly identifies a large number of individual cells but is only able to measure hundreds of genes simultaneously (i.e. $p\ll n$). Each entry $y_{ij}\in\mathbb{N},i=1,\ldots,n,j=1,\ldots,p$ is the read count observed in sample $i$ (with known location) for gene $j$. Let a $n$-by-$2$ matrix $\bm{T}$ denote the related geospatial profile, where each row $\bm{t}_i=(t_{i1},t_{i2})\in\mathbb{R}^2$ gives the coordinates in a compact subset of the two-dimensional Cartesian plane for sample $i$. 



\subsection{Modeling sequence count data via a ZINB model}\label{sec_zinb}
In addition to over-dispersion, sequence count data suffer from zero-inflation, especially when the sequence depth is not enough. For instance, the real data analyzed in \cite{marioni2008rna} and \cite{witten2010ultra} contain $40\%$ -- $50\%$ zeros of all numerical values. Single-cell RNA-seq (scRNA-seq) data usually have a higher proportion of zero read counts, compared with bulk RNA-seq data \citep[e.g.][]{kharchenko2014bayesian,finak2015mast,lun2016pooling}. In general, zero-inflation arises for both biological reasons (e.g. subpopulations of cells or transient states where a gene is not expressed) and technique reasons (e.g. dropouts, where a gene is expressed but not detected through sequencing due to limitation of the sampling effort). The gene expression profiles of the mouse olfactory bulb data and the human breast cancer data analyzed in this paper contains about $60\%$ and $90\%$ zeros, respectively.

Accommodating towards these two characteristics, we start by considering a zero-inflated negative binomial (ZINB) model to model the read counts,
\begin{equation}\label{zinb}
y_{ij}\quad\sim\quad\pi_i\text{I}(y_{ij}=0)+(1-\pi_i)\text{NB}(s_i\lambda_{ij},\phi_j),
\end{equation}
where we use $\text{I}(\cdot)$ to denote the indicator function and $\text{NB}(\mu,\phi),\mu,\phi>0$ to denote a negative binomial (NB) distribution with expectation $\mu$ and dispersion $1/\phi$. Here we constrain one of the two mixture kernels to be degenerate at zero, thereby allowing for zero-inflation. The sample-specific parameter $\pi_i\in(0,1)$ can be viewed as the proportion of extra zero (i.e. false zero or structural zero) counts in sample $i$. With this NB parameterization, the p.m.f. is written as $\frac{\Gamma(y+\phi)}{y!\Gamma(\phi)}\left(\frac{\phi}{\mu+\phi}\right)^\phi\left(\frac{\mu}{\mu+\phi}\right)^y$, with the variance $\text{Var}(Y)=\mu+\mu^2/\phi$, therefore allowing for over-dispersion. A small value of $\phi$ indicates a large variance to mean ratio, while a large value approaching infinity reduces the NB model to a Poisson model with the same mean and variance. 

The NB mean is decomposed of two multiplicative effects, the size factor $s_i$ and the normalized expression level $\lambda_{ij}$. The collection $\bm{s}=(s_1,\ldots,s_n)$ reflects many nuisance effects across samples, including but not limited to: 1) reverse transcription efficiency; 2) amplification and dilution efficiency; 3) sequencing depth. Once the global sample-specific effect is accounted for, $\lambda_{ij}$ can be interpreted as the normalized expression level of gene $j$ observed at sample $i$. Note that such a multiplicative characterization of the NB of Poisson mean is typical in both the frequentist \citep[e.g.][]{Witten2011,Li2012,Cameron2013} and the Bayesian literature \citep[e.g.][]{Banerjee2014,Airoldi2016} to justify latent heterogeneity and extra over-dispersion in multivariate count data. To ensure identifiability between these two classes of parameters, we follow \cite{sun2020statistical} to set $s_i$ proportional to the summation of the total number of read counts across all genes for sample $i$, combined with a constraint of $\prod_{i=1}^ns_i=1$. It results in $s_i=\sum_{j=1}y_{ij}/\prod_{i=1}^n\sum_{j=1}y_{ij}$. If the main interest is in the absolute gene expression level, $s_i$'s can be set to $1$ \citep{li2019bayesian}. Our modeling approach yields a de-noised version of gene expression data, i.e. $\lambda_{ij}$'s, after characterizing zero-inflation (via $\pi_i$'s), over-dispersion (via $\phi_j$'s), and sample heterogeneity (via $s_i$'s).

Now we rewrite model (\ref{zinb}) by introducing a latent indicator variable $\eta_{ij}$, which follows a Bernoulli distribution with parameter $\pi_i$,
\begin{equation}
y_{ij}|\eta_{ij}\quad\sim\quad\begin{cases}
{\begin{array}{ll}
	0 & \text{ if }  \eta_{ij}=1\\
	\text{NB}(y_{ij};s_i\lambda_{ij},\phi_j) & \text{ if } \eta_{ij}=0\\
	\end{array}}
\end{cases}.
\end{equation} 
The independent Bernoulli prior assumption can be further relaxed by formulating a $\text{Be}(a_\pi,b_\pi)$ hyperprior on $\pi_i$, leading to a beta-Bernoulli prior of $\eta_{ij}$ with expectation $a_\pi/(a_\pi+b_\pi)$.  Setting $a_\pi=b_\pi=1$ results in a noninformative prior on $\pi_i$. For all dispersion parameters $\phi_j$'s, we assume a gamma distribution, i.e. $\phi_j\sim\text{Ga}(a_\phi,b_\phi)$. We recommend small values, such as $a_\phi=b_\phi=0.001$, for a noninformative setting \citep{gelman2006prior}.

\subsection{Identifying SV genes via a geostatistical mixture model}\label{sec_gaussian_process}
We incorporate the covariates and spatial random process into the model construction by specifying a geostatistical model \citep{gelfand2016spatial} for the normalized expression level of gene $j$ for sample $i$ at location $\bm{t}_{i}$,
\begin{equation}\label{loglink}
\log\lambda_{ij}\quad=\quad\bm{x}_i^T\bm{\beta}_{j}+w_j(\bm{t}_{i}),
\end{equation}
where $\bm{x}_i$ is a $R$-dimensional column vector of covariates that includes a scalar of one for the intercept and $R-1$ measurable explanatory variables for sample $i$ at location $\bm{t}_{i}$. These explanatory variables could be cell types, tissue microenvironment, cell-cycle information, and other information that might be important to adjust for during the analysis. $\bm{\beta}_{j}$ is a $R$-dimensional column vector of coefficients that includes an intercept representing the mean log-expression of gene $j$ across spatial locations. We denote the collection $(w_j(\bm{t}_{1}),\ldots,w_j(\bm{t}_{n}))^T$ by $\bm{w}_j(\bm{T})$ and assume $\bm{w}_j(\bm{T})$ is a gene-specific zero-mean stationary Gaussian process (GP), modeling the spatial correlation pattern among spatial locations through the covariance $\sigma_j^2\bm{K}(\bm{T})$ in a multivariate normal (MVN) distribution, where $\sigma_j^2$ is a scaling factor and the kernel $\bm{K}(\bm{T})$ is a positive definite matrix with each diagonal entry being one and each off-diagonal entry being a function of the relative position (e.g. Euclidean distance) between each pair of locations, ${k}_{ij}(||\bm{t}_{i}-\bm{t}_{j}||)\in[0,1)$. 

The choice of an appropriate kernel function is of critical importance in spatial modeling. A comprehensive overview of many covariance functions can be found in Chapter 4 of \cite{williams2006gaussian}. The use of white noise kernel $\bm{K}=\bm{I}$ indicates that i.i.d. noise, multiplied by $\sigma_j^2$, is added to the mean log-expression of gene $j$ determined by the covariates. Consequently, no spatial correlation w.r.t. underlying gene expression levels should be observed across the space $\bm{T}$. In contrast, a spatially variable (SV) gene is the one that displays significant spatial expression pattern, which is usually defined by a squared exponential (SE) kernel \citep[e.g.][]{svensson2018spatialde,sun2020statistical}, where $k_{ij}=\exp(-||\bm{t}_{i}-\bm{t}_{j}||^2/2l_j^2)$ with the gene-specific characteristic length-scale denoted by $l_j$. A large value of $l_j$ encourages the underlying expression levels of distant spots or cells become more correlated.  

To identify a subset of gene that are spatially variable across locations, we postulate the existence of a latent binary vector $\bm{\gamma}=(\gamma_1,\ldots,\gamma_p)^T$, with $\gamma_j=1$ if gene $j$ is a SV gene, and $\gamma_j=0$ otherwise. Formulating this assumption, we can rewrite model (\ref{loglink}),
\begin{equation}\label{mixgp}
\log\bm{\lambda}_{j}|\gamma_{j},l_j\quad\sim\quad\begin{cases}
{\begin{array}{ll}
	\text{MVN}(\bm{X}\bm{\beta}_{j},\sigma_j^2\bm{K}) & \text{ if }  \gamma_{j}=1\\
	\text{MVN}(\bm{X}\bm{\beta}_{j},\sigma_j^2\bm{I})  & \text{ if } \gamma_{j}=0\\
	\end{array}}
\end{cases}.
\end{equation}
A common choice for the prior of the binary latent vector $\bm{\gamma}$ is a product of Bernoulli distributions on each individual component with a common hyperparameter $\omega$, i.e. $\gamma_j\sim\text{Bern}(\omega)$. It is equivalent to a binomial prior on the number of SV genes, i.e. $p_\gamma=\sum_{j=1}^p\gamma_j\sim\text{Bin}(p,\omega)$. The hyperparameter $\omega$ can be elicited as the proportion of genes expected \textit{a priori} to be spatially variable. This prior assumption can be further relaxed by formulating a $\text{Be}(a_\omega,b_\omega)$ hyperprior on $\omega$, which leads to a beta-binomial prior on $p_\gamma$ with expectation $pa_\omega/(a_\omega+b_\omega)$. Suggested by \cite{tadesse2005bayesian}, we impose a constraint of $a_\omega+b_\omega=2$ for a vague prior of $\omega$.

Taking a conjugate Bayesian approach, we impose a normal prior on each coefficient in $\bm{\beta}_{j}$ and an inverse-gamma (IG) prior on $\sigma_j^2$; that is, $\bm{\beta}_{j} \sim\text{MN}(\bm{0},h\sigma_j^2\bm{I})$ and $\sigma_{j}^2 \sim\text{IG}(a_\sigma,b_\sigma)$. This parameterization setting is standard in most Bayesian normal models. It allows for creating a computationally efficient feature selection algorithm by integrating out the GP mean and covariance scaling factor. The integration leads to marginal non-standardized multivariate Student's t-distributions (MVT) on $\log\bm{\lambda}_{j}$. Consequently, we can write the collapsed version of model (\ref{mixgp})
\begin{equation}\label{alpha_t}
\begin{split}
&\log\bm{\lambda}_{j}|\gamma_j,{l}_j\quad\sim\\
&\quad\begin{cases}
{\begin{array}{l}
	\text{MVT}_{2a_\sigma}\left(\bm{0},\frac{b_\sigma}{a_\sigma}\left(\bm{K}^{-1}-\bm{K}^{-1}\bm{X}\bm{G}^{-1}{\bm{X}}^T\bm{K}^{-1}\right)^{-1}\right)\\ \hfill \text{ if }   \gamma_j=1\\
	\text{MVT}_{2a_\sigma}\left(\bm{0},\frac{b_\sigma}{a_\sigma}\left(\bm{I}-\bm{X}({\bm{X}}^T\bm{X}+\bm{I}/h)^{-1}{\bm{X}}^T\right)^{-1}\right)\\ \hfill\text{ if }   \gamma_j=0\\
	\end{array}}
\end{cases},
\end{split}
\end{equation}
where $\bm{G}={\bm{X}}^T\bm{K}^{-1}\bm{X}+\bm{I}/h$. To complete the model specification, we choose a uniform prior $l_j\sim\text{U}(a_l,b_l)$. Following \cite{sun2020statistical}, we suggest the value of $a_l$ and $b_l$ set to be $t^\text{min}/2$ and $2t^\text{max}$, where $t^\text{min}$ and $t^\text{max}$ are the minimum and maximum value of the non-zero Euclidean distances across all pairs of spatial locations, respectively.

\section{Model Fitting}\label{model_fitting}
In this section, we briefly describe the Markov chain Monte Carlo (MCMC) algorithm for posterior inference. Our inferential strategy allows us to simultaneously infer the normalized expression level of each gene across different locations, ${\lambda}_{ij}$'s, while identifying the SV genes through ${\gamma}_j$'s.

\subsection{MCMC algorithm}
The model parameter space consists of $(\bm{H},\bm{\phi},\bm{\Lambda},\bm{\gamma},\bm{l})$, where $\bm{H}=\{\eta_{ij},i=1,\ldots,n,j=1,\ldots,p\}$ is the extra zero (i.e. false or structural zero) indicator matrix, $\bm{\phi}=\{\phi_j,j=1,\ldots,p\}$ is the collection of dispersion parameters for all genes, $\bm{\Lambda}=\{{\lambda}_{ij},i=1,\ldots,n,j=1,\ldots,p\}$ is the collection of normalized expression levels, $\bm{\gamma}=\{\gamma_j,j=1,\ldots,p\}$ is the SV gene indicator vector, and $\bm{l}$ is the kernel parameters. We start by writing the full posterior,
\begin{equation}
\begin{split}
\pi(\bm{H},\bm{\phi},\bm{\Lambda},\bm{\gamma},\bm{l}|\bm{Y})\quad\propto\quad & f(\bm{Y}|\bm{H},\bm{\phi},\bm{\Lambda})\pi(\bm{H})\pi(\bm{\phi})\\
& \pi(\bm{\Lambda}|\bm{H},\bm{\gamma},\bm{l})\pi(\bm{l}|\bm{\gamma})\pi(\bm{\gamma}).
\end{split}
\end{equation}
According to Section \ref{sec_zinb}, we can compute the likelihood and priors as
\begin{equation*}
\begin{split}
f(\bm{Y}|\bm{H},\bm{\phi},\bm{\Lambda})=& \prod_{i=1}^n\prod_{\{j:\eta_{ij}=0\}}\text{NB}(y_{ij};s_i\lambda_{ij},\phi_j),\\
\pi(\bm{H})=&\prod_{i=1}^n\prod_{j=1}^p\text{Be-Bern}(\eta_{ij};a_\pi,b_\pi),\\
\pi(\bm{\phi})=&\prod_{j=1}^p\text{Ga}(\phi_j;a_\phi,b_\phi),
\end{split}
\end{equation*}
and according to Section \ref{sec_gaussian_process}, we can calculate the priors as
\begin{equation*}
\begin{split}
\pi(\bm{\Lambda}|\bm{H},\bm{\gamma},\bm{l})=&\prod_{j=1}^p\pi(\log\bm{\lambda}_{j}|\gamma_j,{l}_j),\\
\pi(\bm{\gamma})=&\prod_{j=1}^p\text{Be-Bern}(\gamma_j;a_\omega,b_\omega),\\
\pi(\bm{l}|\bm{\gamma})=&\prod_{\{j:\gamma_j=1\}}\text{Ga}(l_j;a_l,b_l),
\end{split}
\end{equation*}
where the explicit formula of $\pi(\log\bm{\lambda}_{j}|\gamma_j,{l}_j)$ is given in (\ref{alpha_t}).

Identifying SV genes through the selection vectors $\bm{\gamma}$ is our main interest. To serve this purpose, a MCMC algorithm is designed based on Metropolis search variable selection algorithms \citep{George1997,Brown1998}. As discussed in Section \ref{sec_gaussian_process}, we have integrated out the mean and covariance scaling factor. This step helps us speed up the MCMC convergence and improve the estimation of $\bm{\gamma}$. At each MCMC iteration, we perform the following steps:

{\bf Update of zero-inflation indicator $\bm{H}$:} We update each $\bm{\eta}_j,j=1,\ldots,p$ separately. For each gene $j$, we update the hidden zero-inflation indicator for sample $i$ $\eta_{ij},i=1,\ldots,n$ that corresponds to $y_{ij}=0$ separately by sampling from the normalized version of the following conditional:
\begin{equation*}
p(\eta_{ij}|\cdot)\propto \text{NB}(y_{ij};s_i\lambda_{ij},\phi_j)\pi(\log\bm{\lambda}_{j}|\gamma_j,{l}_j)\text{Be-Bern}(\eta_{ij};a_\pi,b_\pi).
\end{equation*}
Note that the second term can be ignored here due to the limited contribution of a single location towards the calculation the whole normalized expression level across the space.

{\bf Update of dispersion parameter $\bm{\phi}$:} We update each ${\phi}_{j},j=1,\ldots,p$ separately by using a random walk Metropolis-Hastings (RWMH) algorithm. We first propose a new ${\phi_{j}}^*$, of which logarithmic value is generated from $\text{N}\left(\log\phi_{j},\tau_\phi^2\right)$ and then accept the proposed value ${\phi_{j}}^*$ with probability $\min(1,m_\text{MH})$, where the Hastings ratio is 
\begin{equation*}
m_\text{MH}=\prod_{\{i:\eta_{ij}=0\}}\frac{\text{NB}(y_{ij};s_i\lambda_{ij},\phi_j^*)}{\text{NB}(y_{ij};s_i\lambda_{ij},\phi_j)}\frac{\text{Ga}(\phi_j^*;a_\phi,b_\phi)}{\text{Ga}(\phi_j;a_\phi,b_\phi)}.
\end{equation*}
Note that the proposal density ratio cancels out for this RWMH update. 

{\bf Update of normalized gene expression levels $\bm{\Lambda}$:} We update each $\bm{\lambda}_j,j=1,\ldots,p$ separately. For each gene $j$, we update its normalized gene expression at location $\bm{t}_{i,\cdot},i=1,\ldots,n$ sequentially by using the RWMH algorithm. We first propose a new $\lambda_{ij}^*$ from $\text{N}\left(\lambda_{ij},\tau_\lambda^2\right)$, and then accept the proposed value with probability $\min(1,\text{m}_\text{MH})$, where the Hastings ratio is 
\begin{equation*}
\text{m}_\text{MH}=\frac{\text{NB}(y_{ij};s_i\lambda_{ij}^*,\phi_j)}{\text{NB}(y_{ij};s_i\lambda_{ij},\phi_j)}\frac{\pi(\log\bm{\lambda}_{j}^*|\gamma_j,{l}_j)}{\pi(\log\bm{\lambda}_{j}|\gamma_j,{l}_j)}.
\end{equation*}
Note that the proposal density ratio cancels out for this RWMH update. If $\eta_{ij}=1$, then the first term also cancels out.

{\bf Joint update of SV gene indicator $\bm{\gamma}$ and kernel parameter $\bm{l}$:} We perform a between-model step to update these two groups of parameters jointly since $\bm{l}$ depends on $\bm{\gamma}$. This is done via an {\it add-delete} algorithm. In this approach, a new candidate vector, say ${\bm{\gamma}}^*$, is generated by randomly choosing an entry of $\bm{\gamma}$, say $j$, and changing its value to $\gamma_j^*=1-\gamma_j$. Then, this proposed move is accepted with probability $\text{min}(1,\text{m}_\text{MH})$, where the Hastings ratio is 
\begin{equation*}
\text{m}_\text{MH}=g(p_\gamma)\frac{\pi(\log\bm{\lambda}_{j}|\gamma_j^*,{l}_j^*)}{\pi(\log\bm{\lambda}_{j}|\gamma_j,{l}_j)}\frac{\text{Ga}(l_j^*;a_l,b_l)^{\gamma_j^*}}{\text{Ga}(l_j;a_l,b_l)^{\gamma_j}}\frac{J\left({l}_j\leftarrow{l}_j^*|{\gamma}_j\leftarrow{{\gamma}_j^*}\right)}{J\left({l}_j^*\leftarrow{l}_j|{\gamma}_j^*\leftarrow{{\gamma}_j}\right)},
\end{equation*}	
where we use $J(\cdot\leftarrow\cdot)$ to denote the proposal probability distribution for the selected move. For the \textit{add} case, we propose a new $l_j^*$, of which logarithmic value is drawn from a truncated normal distribution $\text{N}_{[a_l,b_l]}\left(\log(t^\text{min}/2),(10\tau_l)^2\right)$. The first term is a function with $p_\gamma$ resulted from $\pi(\bm{\gamma}^*)/\pi(\bm{\gamma})$, which equal to $\frac{p_\gamma+a_\pi}{p_\gamma+1}\frac{n-p_\gamma+1}{n-p_\gamma+b_\pi}$ for the \textit{add} step and the reciprocal for the \textit{delete} step. The last proposal density ratio equals to
\begin{equation*}
\begin{cases}
{\begin{array}{ll}
	1/\text{N}(\log l_j^*;\log(t^\text{min}/2),(10\tau_l)^2) & \text{ for } \textit{add}\\
	\text{N}(\log l_j;\log(t^\text{min}/2),(10\tau_l)^2) & \text{ for } \textit{delete}
	\end{array}}
\end{cases}.
\end{equation*}

{\bf Update of kernel parameter $\bm{l}$:} We perform a within-model step to update each ${l}_{j}$ of which $\gamma_j=1$ separately by using the RWMH algorithm. Specifically, we propose a new ${l_{j}}^*$, of which logarithmic value is sampled from a truncated normal distribution $\text{N}_{[a_l,b_l]}\left(\log l_j,\tau_l^2\right)$, and then accept the proposed value ${l_{j}}^*$ with probability $\min(1,m_\text{MH})$, where the Hastings ratio is 
\begin{equation*}
m_\text{MH}=\frac{\pi(\log\bm{\lambda}_{j}|\gamma_j,{l}_j^*)}{\pi(\log\bm{\lambda}_{j}|\gamma_j,{l}_j^*)}.
\end{equation*}
Note that both the prior ratio and proposal density ratio cancel out. 

\subsection{Posterior inference}
Our primary interest lies in the identification of SV genes via the selection vector $\bm{\gamma}$. One way to summarize the posterior distribution of the parameters of interest is via \textit{maximum-a-posteriori} (MAP) estimates. 
\begin{equation}
\begin{split}
\bm{\gamma}^\text{MAP}\quad&=\quad\text{argmax}_{\bm{\gamma}}\pi(\bm{\gamma}|\bm{H},\bm{\phi},\bm{\Lambda},\bm{l},\bm{Y})\\
\quad&=\quad\text{argmax}_{\bm{\gamma}}f(\bm{Y}|\bm{H},\bm{\phi},\bm{\Lambda})\pi(\bm{\Lambda}|\bm{H},\bm{\gamma},\bm{l})\pi(\bm{\gamma}).
\end{split}
\end{equation}
A more comprehensive summarization of $\bm{\gamma}$ is to select the SV gene based on their marginal distributions. In particular, we estimate marginal posterior probabilities of inclusion (PPI) of each single gene by $\text{PPI}_j=\sum_{u=1}^U\left( \gamma_j\text{ at iteration }u\right) / U$, where $U$ is the total number of iterations after burn-in. The marginal PPI represents the proportion of MCMC samples in which the gene is selected to be spatially variable. A set of SV genes can be picked based on their PPIs. 
\begin{equation}
\bm{\gamma}^\text{PPI}\quad=\quad \left(\text{I}(\text{PPI}_1\ge c),\ldots,\text{I}(\text{PPI}_p\ge c)\right)^T
\end{equation}
For instance, the selection can be done by including those genes with marginal PPIs greater than a pre-specified value such as $c=0.5$, also known as the median model. Alternatively, we can choose the threshold that controls for multiplicity \citep{Newton2004}, which guarantees the expected Bayesian false discovery rate (FDR) to be smaller than a number. The Bayesian FDR is calculated as follows,
\begin{align}\label{BFDR}
\text{FDR}(c)\quad=\quad\frac{\sum_{j=1}^{p}(1-\text{PPI}_j)\text{I}(1-\text{PPI}_j<c)}{\sum_{j=1}^{p}\text{I}(1-\text{PPI}_j<c)}.
\end{align}
Here $\text{FDR}(c)$ is the desired significance level and $c$ is the corresponding threshold on the PPI, with $\text{FDR}(c)=0.05$ being generally used in other parametric test settings for the spatial transcriptomics studies \citep[see e.g.][]{edsgard2018identification,svensson2018spatialde,sun2020statistical}.

In addition to using PPI as a significance measurement, we could obtain a $p$-value using likelihood test, where the Bayes factor could be approximated by averaging the ratios of the two probability densities in Equation \ref{alpha_t} over all iterations after burn-in. The likelihood ratio asymptotically follows a chi-square distribution with degree of freedom being $1$ under null hypothesis.

\section{Simulation Study}\label{simulation}
\begin{figure}[!h]
	\centerline{\includegraphics[width=1\linewidth]{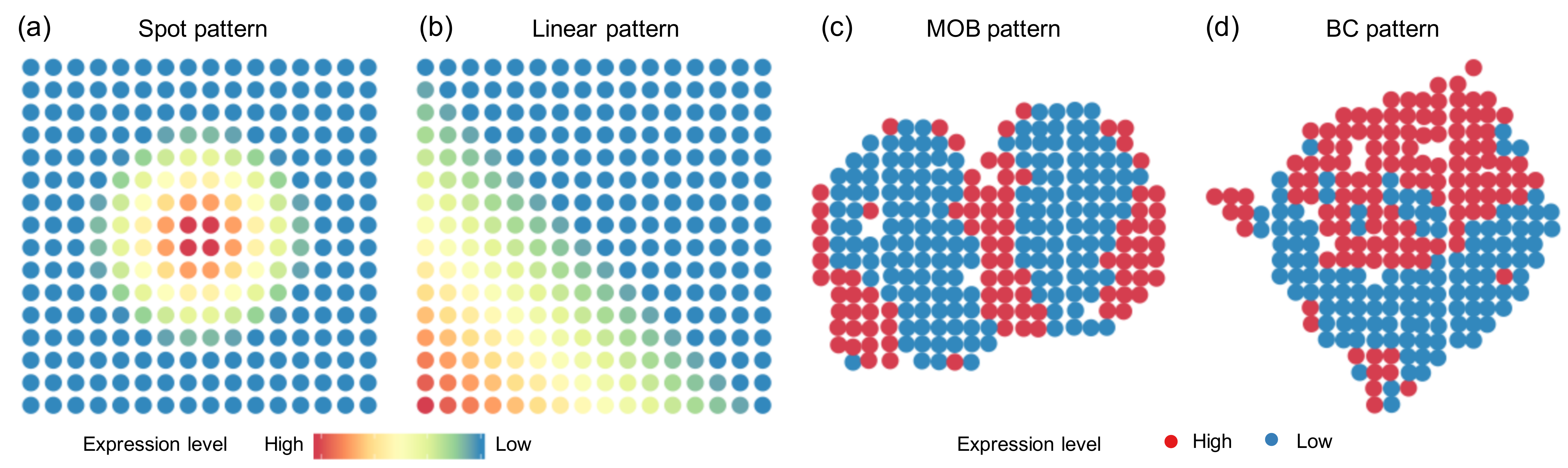}}
	\caption{The four spatial patterns used in the simulation study: (a) and (b) Two artificially generated patterns; (c) and (d) Two binary patterns that were summarized on the basis of spatially variable (SV) genes identified by SPARK in mouse olfactory bulb (MOB) and human breast cancer (BC) data, respectively.}\label{Figure 1}
\end{figure}

We used synthetic data to assess the performance of BOOST-GP and compare it with that of the only three available methods, i.e. Trendsceek \citep{svensson2018spatialde}, SpatialDE \citep{edsgard2018identification}, and SPARK \citep{sun2020statistical}. We followed the data generative schemes originated by \cite{edsgard2018identification} and \cite{sun2020statistical} based on two artificially generated spatial patterns and two real spatial patterns in mouse olfactory bulb (MOB) and human breast cancer (BC) data. The first two namely spot and linear patterns, which are shown in Figure \ref{Figure 1}(a) and (b), respectively, were on a $16$-by$16$ square lattice ($n=256$ spots). The MOB and BC patterns, which are shown in Figure \ref{Figure 1}(c) and (d), respectively, were on $n=260$ and $250$ spots collected in the MOB study replicate 11 and BC study layer 2, after filtering those spots less than $10$ total read counts. We simulated $p=100$ genes, among which $15$ were SV genes while the rest were non-SV genes. For each gene $j$ at each spot $i$, its latent normalized expression level on a logarithmic scale was the sum of three components, \[\log(\lambda_{ij})\quad=\quad\beta_j+e_i+\epsilon_{ij},\] where $\beta_j$ is the gene-specific baseline, $e_i$ is the spot-specific fold-change between SV and non-SV genes, and $\epsilon_{ij}$ is the non-spatial random error. In our simulation study, we set the baseline $\beta_j=2$ and assumed $\epsilon_{ij}\sim\text{N}(0,0.3^2)$. The spatial patterns were embedded in the construction of $e_i$'s. For a non-SV gene, we set $e_i=0,\forall i$ so that the latent normalized expression levels were independent and identically distributed from a log-normal distribution with mean and variance being $2$ and $0.3^2$. Consequently, no spatial correlation should be observed. For an SV gene with the spot pattern, the values of $e_i$'s corresponding to the four center spots were set to $\log6$, while all others were linearly decreased to $0$ within the radius of $5$. For an SV gene with the linear pattern, the value of $e_i$ corresponding to the most bottom-left spot was set to $\log6$, while all others were linearly decreased to $0$ along the $45^\circ$ line. For an SV gene with the real patterns, each spot was categorized into two groups. We set $e_i=0$ for those spots from the low expression group, while altering its value to $\log3$ for those spots from the high expression group. The difference in $e_i$'s between the two groups of spots thus introduced spatial differential expression patterns. Then, we simulated each gene expression count data $y_{ij}$ from an NB distribution where its mean was a product of the latent normalized expression level $\lambda_{ij}$ and the size factor $s_i$. We sampled $s_i$'s from a log-normal distribution with mean zero and variance $0.2^2$. The gene-specific NB dispersion parameters were sampled from an exponential distribution with mean $10$. Furthermore, to mimic the excess zeros observed in the real data, we randomly chose $30\%$ spots and forced their counts to be zero. Combined with the four patterns (i.e. spot, linear, MOB, and BC) and two count generating models (i.e. NB and ZINB with $30\%$ false zeros), there were $4\times2=8$ scenarios in total. For each of the scenarios, we independently repeated the above steps to generate $10$ datasets.

For prior specification of the proposed Bayesian model, we recommended and used the following default settings. The hyperparameters that controlled the percentage of extra zeros \textit{a priori} were set to $\pi\sim\text{Be}(a_\pi=1, b_\pi=1)$. As for the gamma priors on the NB dispersion and GP characteristic length-scale parameters, i.e. $\phi_j\sim\text{Ga}(a_\phi,b_\phi)$ and $l_j\sim\text{Ga}(a_l,b_l)$, we set $a_\phi$, $b_\phi$, $a_l$ and $b_l$ to $0.001$, which led to a vague distribution with mean and variance equal to $1$ and $1,000$. We set the hyperparameters that control the selection of SV genes, $\omega\sim\text{Be}(a_\omega=0.1,b_\omega=1.9)$, resulting in the proportion of SV genes expected \textit{a priori} to be $a_\omega/(a_\omega+b_\omega)=5\%$. As for the inverse-gamma priors on the variance components $\sigma_j^2$, we set the shape parameters $a_\sigma=3$ and the scale parameters $b_\sigma=1$ to achieve a fairly flat distribution with an infinite variance. We further set the default values of $h$ to $10$. As for the BOOST-GP algorithm setting implemented in this paper, we ran four independent MCMC chain with $2,000$ iterations, discarding the first $50\%$ sweeps as burn-in. We started each chain from a model by setting all genes to be non-SV and randomly drawing all other model parameters from their prior distribution. Results we report below were obtained by pooling together the MCMC outputs from the four chains. All experiments were implemented in \texttt{R} with \texttt{Rcpp} package to accelerate computations.

To quantify the accuracy of identifying SV genes via the binary vector $\bm{\gamma}$, we consider two widely used measures of the quality of binary classifiers: 1) area under the curve (AUC) of the receiver operating characteristic (ROC); and 2) Matthews correlation coefficient (MCC) \citep{matthews1975comparison}. The former considers both true positive (TP) and false positive (FP) rates across various threshold settings, while the latter balances TP, FP, true negative (TN), and false negative (FN) counts even if the true zeros and ones in $\bm{\gamma}$ are of very different sizes. MCC is defined as \[\text{MCC}\quad=\quad\frac{(\text{TP}\times \text{TN} - \text{FP}\times \text{FN})}{\sqrt{(\text{TP+FP})(\text{TP+FN})(\text{TN+FP})(\text{TN+FN})}}.\] The number of SV genes are usually assumed to be a small fraction of the total. Therefore, MCC is more appropriate to handle such an imbalanced scenario. Note that the AUC yields a value between $0$ to $1$ that is averaged by all possible thresholds used to select discriminatory features based on PPIs or $p$-values, and the MCC value ranges from $-1$ to $1$ to pinpoint a specified threshold. The larger the index, the more accurate the inference.

\begin{figure}[!h]
	\centerline{\includegraphics[width=1\linewidth]{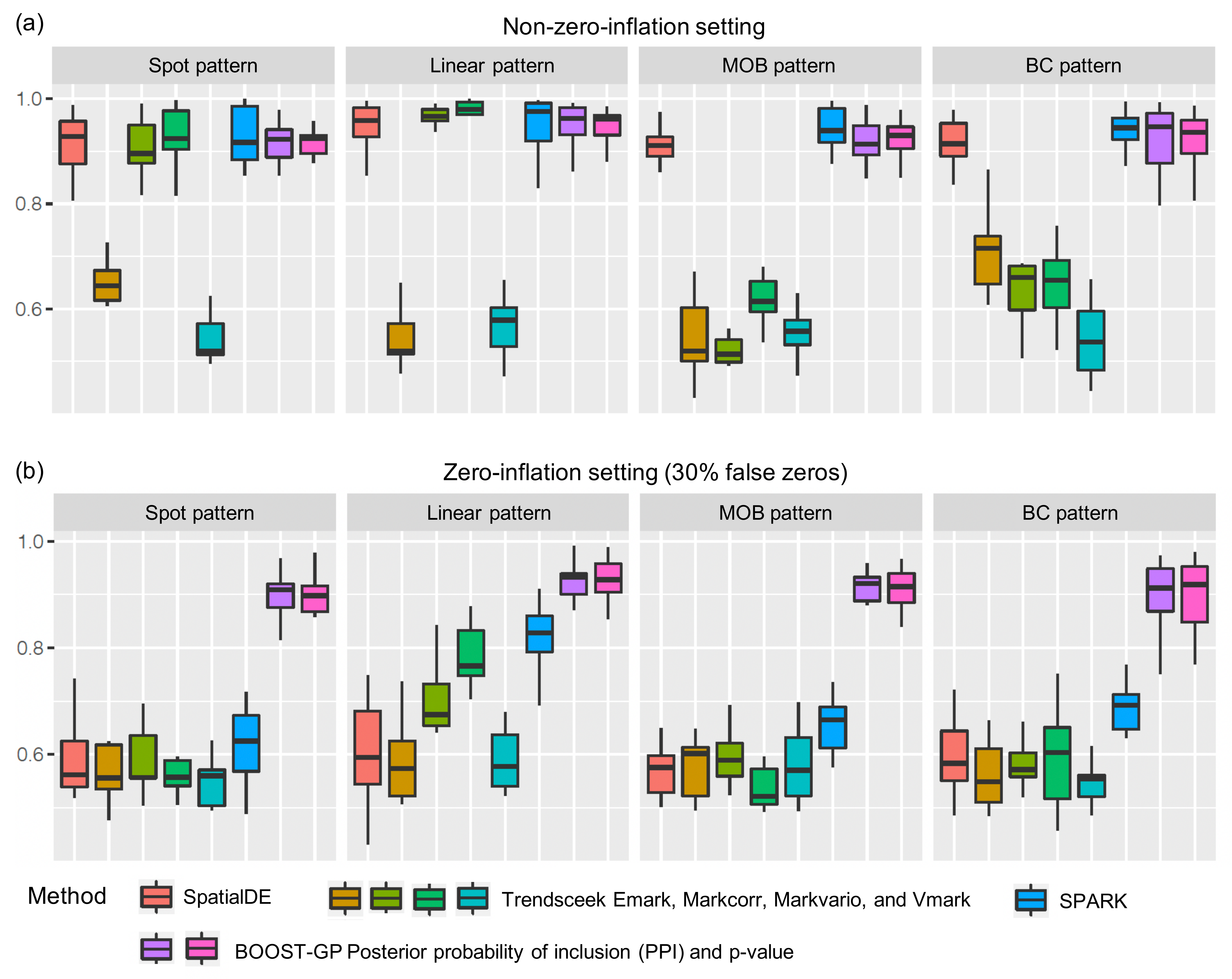}}
	\caption{Simulation study: The boxplots of AUCs achieved by BOOST-GP, SPARK, SpatialDE, and Trendsceek under different scenarios in terms of spatially variable (SV) gene spatial pattern and count generating process}\label{Figure 2}
\end{figure}

To demonstrate the superiority of BOOST-GP, we compared ours with other existing approaches for SV gene detection. Figure \ref{Figure 2} displays the average AUCs by different methods over $10$ replicated datasets under each scenario. We can see that BOOST-GP, SPARK, and SpatialDE had similar power when there were no false zeros in the simulated data. The Markcorr and Markvario of Trendsceek also had close performance in the spot and linear patterns, but performed worse in the other two patterns. The Emark and Vmark of Trendsceek did not generated satisfactory results among any of the four patterns. Because BOOST-GP, SPARK, and SpatialDE are all GP-based methods, it was not surprising that they had similar performance in the non-zero-inflation scenarios. While under the zero-inflation setting, BOOST-GP clearly stood out and still maintained an almost unaffected performance. Meanwhile, all other methods suffered from reduced power, indicating the variance brought by an excess of zero count not being properly resolved.
We further evaluated all methods in terms of MCC. To control for the rate of type-I errors, we adjusted $p$-values from SPARK, SpatialDE, and Trendsceek using the Benjamini-Hochberg (BH) method \citep{benjamini1995controlling} and chose a significance level of $0.05$ to select SV genes. For BOOST-GP, because we have multiple significance measurements, we combined the rules of both BH-adjusted $p$-value smaller than $0.05$ and PPI greater than $0.5$ as selection criteria for SV genes.  Table \ref{Table 1} shows the averaged MCCs of all methods in non-zero-inflation and zero-inflation settings. BOOST-GP had very stable performance among all the four patterns. The performance of SpatialDE and SPARK were good in some patterns but dropped in the others. Trendsceek was not able to report any SV genes after adjustment. This was probably due to the $p$-value calculation procedure of Trendsceek, which highly depends on the permutation times. Under the zero-inflated settings, SpatialDE and SPARK failed the task too in most of the patterns. BOOST-GP still had reasonable performance though the accuracy also dropped.

\begin{table}[!t]
	\centering
	\caption{Simulation study: The averaged MCCs, with standard deviations in parentheses, achieved by BOOST-GP, SPARK, and SpatialDE under different scenarios in terms of spatially variable (SV) gene spatial pattern and count generating process.\label{Table 1}} {
		\begin{tabular}{@{}llllll@{}}\toprule 
			& \multicolumn{4}{c}{Non-zero-inflation setting} \\\cline{2-5} 
			& Spot pattern & Linear pattern & MOB pattern & BC pattern\\\midrule
			SpatialDE & $0.643$ ($0.104$) & $\bm{0.810}$ ($0.123$)  & $0.213$ ($0.163$) & $0.416$ ($0.219$) \\
			SPARK & $\bm{0.770}$ ($0.121$) & $0.425$ ($0.070$)  & $\bm{0.796}$ ($0.061$) & $0.595$ ($0.054$) \\
			BOOST-GP & $0.686$ ($0.111$) & $0.801$ ($0.121$)  & $0.698$ ($0.090$) & $\bm{0.723}$ ($0.073$)\\\midrule
			& \multicolumn{4}{c}{Zero-inflation setting (i.e. $30\%$ false zeros)} \\\cline{2-5} 
			& Spot pattern & Linear pattern & MOB pattern & BC pattern\\\midrule
			SpatialDE & $0.000$ ($0.000$) & $0.000$ ($0.000$)  & $0.000$ ($0.000$) & $0.000$ ($0.000$)\\
			SPARK & $0.062$ ($0.103$) & $0.347$ ($0.155$)  & $0.062$ ($0.103$) & $0.044$ ($0.114$) \\
			BOOST-GP & $\bm{0.237}$ ($0.145$) & $\bm{0.543}$ ($0.073$)  & $\bm{0.208}$ ($0.150$) & $\bm{0.206}$ ($0.192$)\\\bottomrule
	\end{tabular}}{}
\end{table}

\section{Real Data Analysis}\label{real_data}
After validating BOOST-GP in simulation studies, we further applied the method in two real spatial transcriptomics data, the mouse olfactory bulb (MOB) and human breast cancer (BC) data. We applied the same preprocessing steps suggested by \cite{sun2020statistical} to remove non-informative genes and spots (only spots with at least ten total count and genes which had non-zero count in at least $10\%$ of spots were kept). After filtering, the MOB data had $p=11,274$ genes and $n=260$ spots, while the BC data had $p=5,262$ genes and $n=250$ spots. We applied the same prior specification, algorithm setting, and significance criteria that we used in simulation. Because of the poor performance of Trendsceek in simulation studies, we did not include it in real data analysis

\begin{figure}[!h]
	\centerline{\includegraphics[width=1\linewidth]{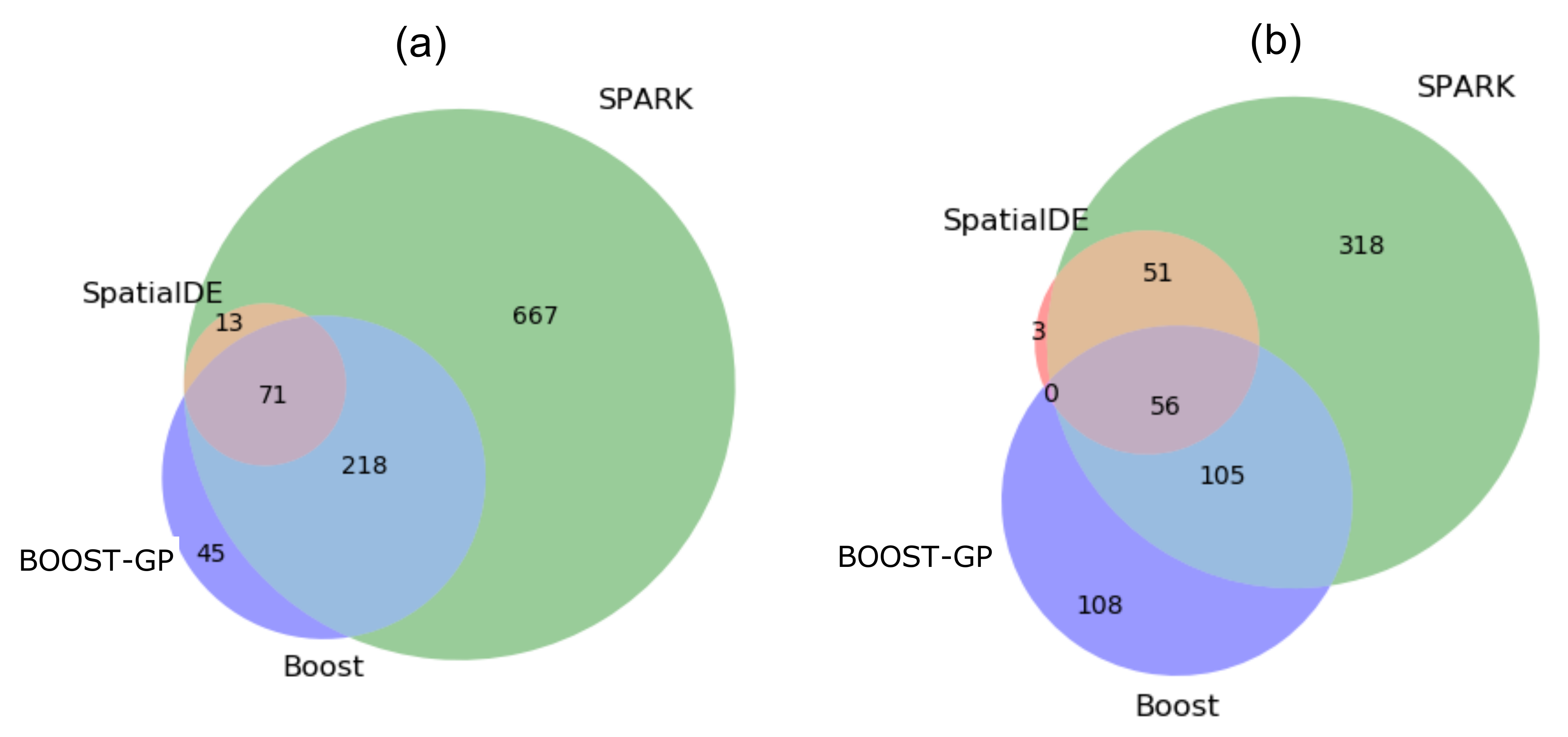}}
	\caption{Real data analysis: The Venn diagram of spatially variable (SV) genes identified by BOOST-GP, SPARK, and SpatialDE in (a) mouse olfactory bulb (MOB) and (b) human breast cancer (BC) data.}\label{Figure 3}
\end{figure}

Figure \ref{Figure 3} shows the overlap of identified SV genes by SpatialDE, SPARK and BOOST- GP methods in MOB and BC datasets. SPARK was shown to be the most aggressive method that reported the most SV genes, while SpatialDE was the most conservative one and reported the fewest SV genes. BOOST-GP lay in between SpatialDE and SPARK, with a lot in common with them, but also uniquely identified genes. Because SpatialDE was almost always a subset of SPARK, we focused on the comparison between BOOST-GP and SPARK in the following analysis. We then did hierarchical clustering on the SV genes identified by BOOST-GP and SPARK, and plotted the averaged normalized expression levels within each cluster to represent each pattern.	

\begin{figure*}[!h]
	\centerline{\includegraphics[width=1\linewidth]{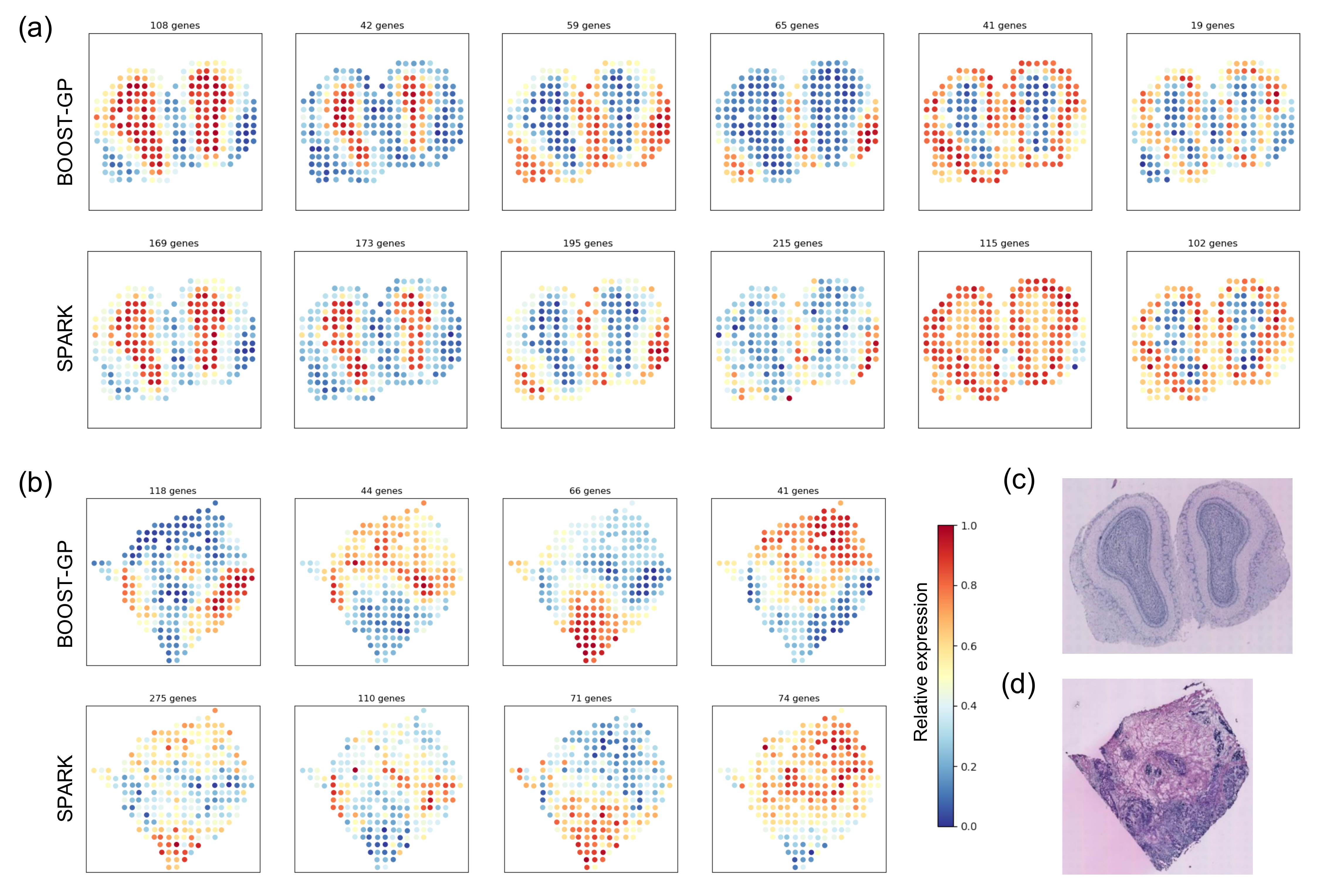}}
	\caption{Real data analysis: (a) and (b) The averaged normalized gene expression levels of spatially variable (SV) genes identified by BOOST-GP and SPARK in (a) mouse olfactory bulb (MOB) and (b) human breast cancer (BC) data, respectively. (c) and (d) The associated hematoxylin and eosin (H\&E)-stained tissue slides of MOB and BC data, respectively.}\label{Figure 4}
\end{figure*}

In order to have a more detailed comparison and not overshadow any subtle patterns, we clustered those SV genes of MOB, as shown in Figure \ref{Figure 4}(a) and (b), into six and four groups. We could see that in MOB data, the patterns detected by BOOST-GP and SPARK resemble each other well. More specifically, the first to sixth patterns can be further merged, leading to three major patterns, which was consistent with what was reported by \cite{sun2020statistical}. This suggested that although BOOST-GP detected fewer SV genes than SPARK, it did not miss any major patterns. In BC data, the patterns summarized by the two methods showed more distinctiveness, especially the first pattern. The remaining three patterns could still largely overlap each other, though not perfectly. This agreed with the Venn diagram that more discrepancy was found between BOOST-GP and SPARK in BC data compared to the MOB data.

\begin{figure*}[!h]
	\centerline{\includegraphics[width=1\linewidth]{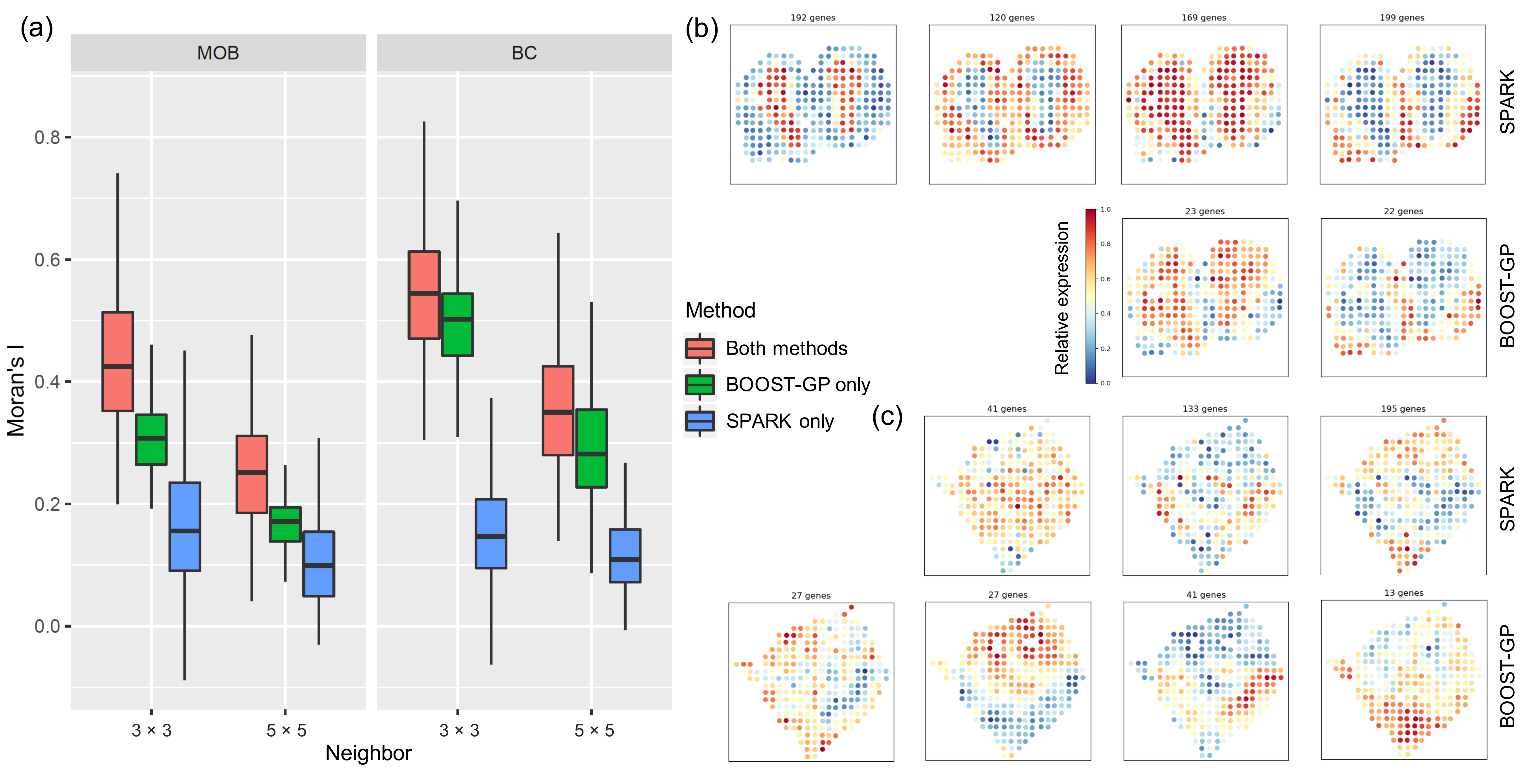}}
	\caption{Real data analysis: (a) The boxplots of Moran's $I$'s of spatially variable (SV) genes identified by both BOOST-GP and SPARK, BOOST-GP only, and SPARK only. (b) and (c) The averaged normalized gene expression levels of spatially variable (SV) genes identified by BOOST-GP only and SPARK only in mouse olfactory bulb (MOB) and human breast cancer (BC) data, respectively.}\label{Figure 5}
\end{figure*}

To have an additional diagnosis of the spatial signals of the detected SV genes, we employed Moran’s $I$ spatial autocorrelation test \citep{moran1950notes,li2007beyond}. The final normalized Moran’s $I$'s were calculated as their difference. The metric ranges from $-1$ to $1$ denoting perfect spatial correlation to complete dispersion, with zero indicating spatial randomness. denoting perfect spatial correlation to complete dispersion, with zero indicating spatial randomness. Figure \ref{Figure 5}(a) summarized the Moran’s $I$'s of the common SV genes found by both methods and SV genes found by individual method alone. Spatial autocorrelation was shown to be the strongest in the common SV gene set, followed by the BOOST-GP only set and then the SPARK only set. The result implies: 1) SV genes identified by both methods are more likely to be true SV genes, and 2) SPARK were in general more aggressive to report more candidates but also with lower average signals. In parallel, we also performed hierarchical clustering on these three sets of genes. In MOB data, $680$ genes identified by SPARK only could be clustered into four groups, as shown in Figure \ref{Figure 5}(b), same as the ones previously described in Figure \ref{Figure 4}(a), while the $45$ genes identified by BOOST-GP only mainly belonged to two patterns, as shown in Figure \ref{Figure 5}(b). This result suggested SPARK might be more sensitive in MOB data, as it is able to find some weak signal genes, but in the meantime, according to Figure \ref{Figure 5}(a), some of those could potentially be false positives that lead to the dilution of Moran’s $I$'s.  In BC data, $372$ genes identified by SPARK only showed both much weaker spatial signals and less spatial patterns compared to $108$ genes identified by BOOST-GP only, as shown in Figure \ref{Figure 5}(c), suggesting the SPARK method potentially scarified specificity while it did not capture true positives as well.

\begin{figure}[!h]
	\centerline{\includegraphics[width=1\linewidth]{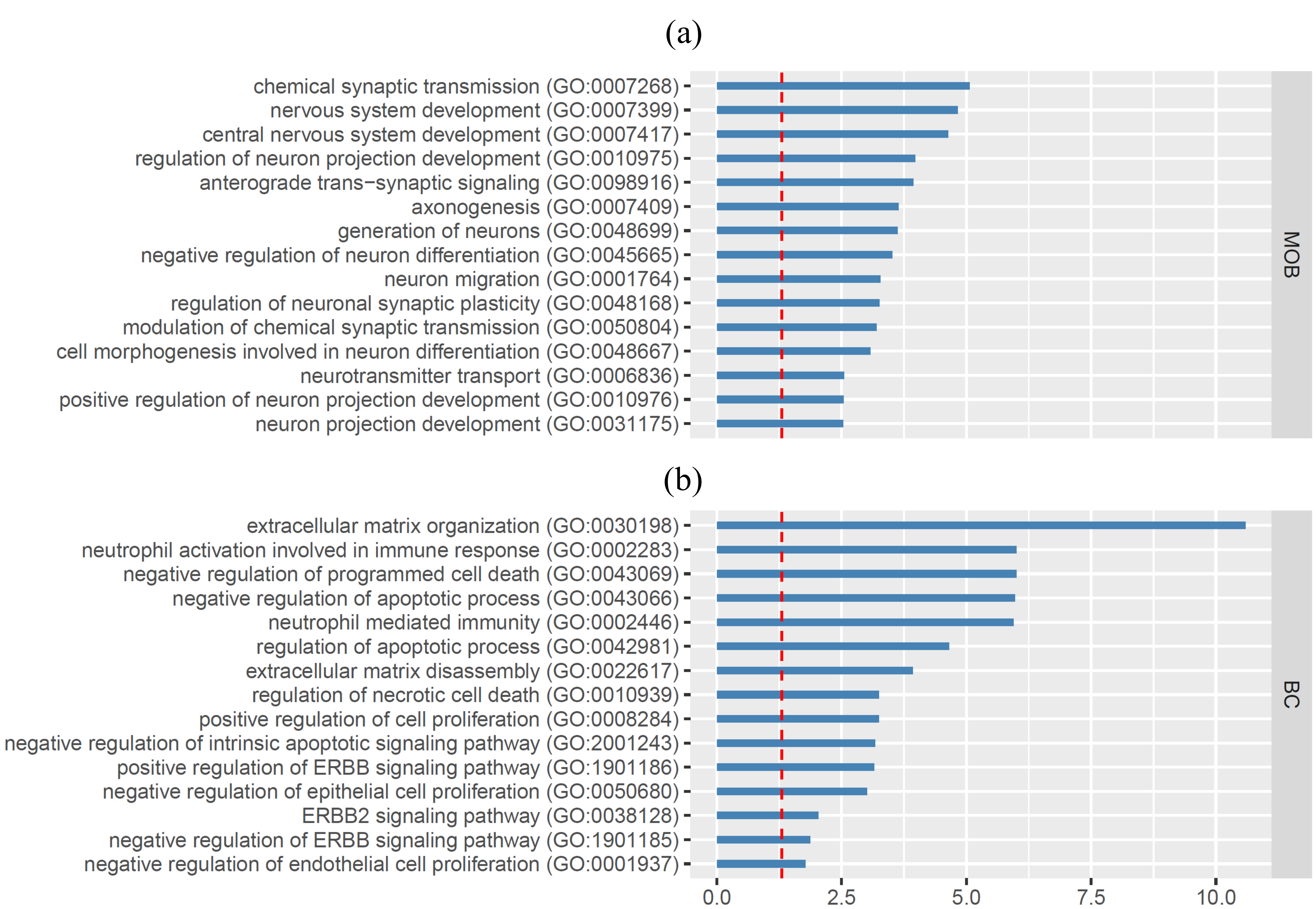}}
	\caption{Real data analysis: Gene ontology (GO) enrichment analysis of spatially variable (SV) genes identified by BOOST-GP in (a) mouse olfactory bulb (MOB) and (b) human breast cancer (BC) data. Red dashed line indicates a significant level of $0.05$.}\label{Figure 6}
\end{figure}

Last, we conducted gene ontology (GO) enrichment analysis to see whether those detected SV genes by BOOST-GP were related to certain biological functions. We performed this on $2,941$ mouse and $2,338$ human GO terms of biological processes which had at least one gene overlap with detected SV gene pools.  Using the adjusted $p$-value smaller than $0.05$ as the threshold, we found $148$ and $197$ enriched biological process GO terms in MOB and BC datasets. We took a closer look at those terms that were associated with the synaptic and nervous system for MOB data, as they play important roles in synaptic organization and nerve development, which is expected be involved in olfactory functioning. As showed in Figure \ref{Figure 6}, many of those synaptic- and nervous-related GO terms were significantly enriched, suggesting the functions of those detected SV genes were aligned with the underlying mechanism that drove the spatially differentiated expression pattern. This was consistent with what was reported in \cite{sun2020statistical}. For BC data, the terms related to extracellular matrix and immune response, as expected, were found to be significantly enriched in the identified SV genes, as they are over-activated in breast cancer. Furthermore, five ERBB-related pathways were found to be significant in our analysis. ERBB family genes, especially ERBB2, are well-known oncogenes in breast cancer that have a close relationship with altered signaling pathways and tumorigenesis and development \citep{stern2000tyrosine}. Three of them were displayed in Figure \ref{Figure 6}. However, none of them were found significant in the SPARK analysis result (e.g. the adjusted $p$-value for ERBB2 signaling pathway (GO:0038128) was $0.239$). We went back and had a deeper check. We found the reason was some genes involved in ERBB pathway were significant in BOOST-GP result but only marginally significant in SPARK (e.g. the adjusted p value and PPI of gene UBB was $9.22\times10^{16}$ and $0.650$ reported by BOOST-GP, the corresponding adjusted p-value for SPARK and SpatialDE were $0.053$ and $0.857$).  It echoed with the previous clustering and autocorrelation analysis, where BOOST-GP had more reliable results in BC data than SPARK.

\section{Conclusion}\label{conclusion}
Recently, the emerging spatial molecular profiling technologies demonstrated great potential in revealing tissue spatial organization and heterogeneity. In this study, we developed a novel Bayesian hierarchical model that incorporates a Gaussian process model to identify spatially variable genes. It can also characterize the embedded temporal or spatial patterns in high-resolution time-series RNA-Seq data \citep[e.g.][]{owens2016measuring} or three-dimensional SMP data \citep[e.g.][]{shah2016situ}. Compared to other existing methods, it has several advantages. First, it directly models count data with a negative binomial distribution that could accommodate the over-dispersion compared to the Poisson distribution. Second and most importantly, it properly addresses the excess of zero count problem, which is typically observed in the SMP data, using a zero-inflation model. Third, it improves the estimation of the kernel length-scale parameter using a Bayesian approach, leading to more stable and accurate results. In simulation studies, BOOST-GP had similar power in identifying spatially variable genes compared with SpatialDE and SPARK, and much better than Trendsceek in general spatial settings. It greatly outperformed all other methods when extra zeros were present in the data. In real data, BOOST-GP and SPARK both had reasonable performance and had a lot in common, while SpatialDE was a bit conservative. The performance of different methods may vary in different datasets with different inherent spatial patterns and data structures. SPARK seemed to be more sensitive in mouse olfactory bulb data with some sacrifice in precision. BOOST-GP performed better in human breast cancer data, identifying undiscovered spatial patterns and meaningful enriched pathways. In general, SPARK was a more aggressive method under the selected significance cutoff, but at the same time bearing a higher risk of discovering more false positives. Because BOOST-GP reports both posterior probability of inclusion and $p$-value as measures of significance, users have more freedom to balance the sensitivity and specificity on their own considerations depending on the study needs. Furthermore, with a Bayesian framework, BOOST-GP can be further extended to incorporate pathway information as prior knowledge to take into account the regulatory relationships between genes to perform a joint estimation.

\section*{Acknowledgements}
The authors would like to thank Jessie Norris for helping us in proofreading the manuscript.

\section*{Funding}
This work has been supported by the National Institutes of Health (NIH) [R35GM136375, P30CA142543, 5R01CA152301, P50CA70907 and 1R01GM115473], and the Cancer Prevention and Research Institute of Texas (CPRIT) [RP190107 and RP180805].
	
	\bibliographystyle{agsm}
    {\bibliography{document.bib}}

@article{benjamini1995controlling,
	title={Controlling the false discovery rate: {A} practical and powerful approach to multiple testing},
	author={Benjamini, Yoav and Hochberg, Yosef},
	journal={Journal of the Royal Statistical Society: Series B (Statistical Methodology)},
	volume={57},
	number={1},
	pages={289--300},
	year={1995},
	publisher={Wiley Online Library}
}

@article{matthews1975comparison,
	title={Comparison of the predicted and observed secondary structure of {T4} phage lysozyme},
	author={Matthews, Brian W},
	journal={Biochimica et Biophysica Acta},
	volume={405},
	number={2},
	pages={442--451},
	year={1975},
	publisher={Elsevier}
}

@article{owens2016measuring,
	title={Measuring absolute {RNA} copy numbers at high temporal resolution reveals transcriptome kinetics in development},
	author={Owens, Nick DL and Blitz, Ira L and Lane, Maura A and Patrushev, Ilya and Overton, John D and Gilchrist, Michael J and Cho, Ken WY and Khokha, Mustafa K},
	journal={Cell reports},
	volume={14},
	number={3},
	pages={632--647},
	year={2016},
	publisher={Elsevier}
}

@article{shah2016situ,
	title={In situ transcription profiling of single cells reveals spatial organization of cells in the mouse hippocampus},
	author={Shah, Sheel and Lubeck, Eric and Zhou, Wen and Cai, Long},
	journal={Neuron},
	volume={92},
	number={2},
	pages={342--357},
	year={2016},
	publisher={Elsevier}
}

@article{moffitt2018molecular,
	title={Molecular, spatial, and functional single-cell profiling of the hypothalamic preoptic region},
	author={Moffitt, Jeffrey R and Bambah-Mukku, Dhananjay and Eichhorn, Stephen W and Vaughn, Eric and Shekhar, Karthik and Perez, Julio D and Rubinstein, Nimrod D and Hao, Junjie and Regev, Aviv and Dulac, Catherine and others},
	journal={Science},
	volume={362},
	number={6416},
	year={2018},
	publisher={American Association for the Advancement of Science}
}

@article{gelfand2016spatial,
	title={Spatial statistics and {G}aussian processes: {A} beautiful marriage},
	author={Gelfand, Alan E and Schliep, Erin M},
	journal={Spatial Statistics},
	volume={18},
	pages={86--104},
	year={2016},
	publisher={Elsevier}
}

@article{tadesse2005bayesian,
	title={Bayesian variable selection in clustering high-dimensional data},
	author={Tadesse, Mahlet G and Sha, Naijun and Vannucci, Marina},
	journal={Journal of the American Statistical Association},
	volume={100},
	number={470},
	pages={602--617},
	year={2005},
	publisher={Taylor \& Francis}
}

@article{gelman2006prior,
	title={Prior distributions for variance parameters in hierarchical models (comment on article by {Browne} and {Draper})},
	author={Gelman, Andrew and others},
	journal={Bayesian Analysis},
	volume={1},
	number={3},
	pages={515--534},
	year={2006},
	publisher={International Society for Bayesian Analysis}
}

@article{Brown1998,
	title={Multivariate {B}ayesian variable selection and prediction},
	author={Brown, Philip J and Vannucci, Marina and Fearn, Tom},
	journal={Journal of the Royal Statistical Society: Series B (Statistical Methodology)},
	volume={60},
	number={3},
	pages={627--641},
	year={1998},
	publisher={Wiley Online Library}
}

@article{George1997,
	title={Approaches for {B}ayesian variable selection},
	author={George, Edward I and McCulloch, Robert E},
	journal={Statistica Sinica},
	pages={339--373},
	year={1997},
	publisher={JSTOR}
}

@article{Newton2004,
	title={Detecting differential gene expression with a semiparametric hierarchical mixture method},
	author={Newton, Michael A and Noueiry, Amine and Sarkar, Deepayan and Ahlquist, Paul},
	journal={Biostatistics},
	volume={5},
	number={2},
	pages={155--176},
	year={2004},
	publisher={Biometrika Trust}
}

@book{williams2006gaussian,
	title={Gaussian processes for machine learning},
	author={Williams, Christopher KI and Rasmussen, Carl Edward},
	volume={2},
	number={3},
	year={2006},
	publisher={MIT press Cambridge, MA}
}

@article{marioni2008rna,
	title={{RNA}-seq: {An} assessment of technical reproducibility and comparison with gene expression arrays},
	author={Marioni, John C and Mason, Christopher E and Mane, Shrikant M and Stephens, Matthew and Gilad, Yoav},
	journal={Genome Research},
	volume={18},
	number={9},
	pages={1509--1517},
	year={2008},
	publisher={Cold Spring Harbor Lab}
}

@article{witten2010ultra,
	title={Ultra-high throughput sequencing-based small {RNA} discovery and discrete statistical biomarker analysis in a collection of cervical tumours and matched controls},
	author={Witten, Daniela and Tibshirani, Robert and Gu, Sam Guoping and Fire, Andrew and Lui, Weng-Onn},
	journal={BMC Biology},
	volume={8},
	number={1},
	pages={58},
	year={2010},
	publisher={Springer}
}

@article{Witten2011,
	title={Classification and clustering of sequencing data using a {Poisson} model},
	author={Witten, Daniela M},
	journal={Ann. Appl. Stat.},
	pages={2493--2518},
	year={2011},
	publisher={JSTOR}
}

@article{Li2012,
	title={Normalization, testing, and false discovery rate estimation for {RNA}-sequencing data},
	author={Li, Jun and Witten, Daniela M and Johnstone, Iain M and Tibshirani, Robert},
	journal={Biostatistics},
	volume={13},
	number={3},
	pages={523--538},
	year={2012},
	publisher={Oxford University Press}
}

@book{Cameron2013,
	title={Regression analysis of count data},
	author={Cameron, A Colin and Trivedi, Pravin K},
	volume={53},
	year={2013},
	publisher={Cambridge University Press}
}

@book{Banerjee2014,
	title={Hierarchical modeling and analysis for spatial data},
	author={Banerjee, Sudipto and Carlin, Bradley P and Gelfand, Alan E},
	year={2014},
	publisher={CRC Press}
}

@article{Airoldi2016,
	title={Improving and evaluating topic models and other models of text},
	author={Airoldi, Edoardo M and Bischof, Jonathan M},
	journal={Journal of the American Statistical Association},
	volume={111},
	number={516},
	pages={1381--1403},
	year={2016},
	publisher={Taylor \& Francis}
}

@article{kharchenko2014bayesian,
	title={Bayesian approach to single-cell differential expression analysis},
	author={Kharchenko, Peter V and Silberstein, Lev and Scadden, David T},
	journal={Nature Methods},
	volume={11},
	number={7},
	pages={740},
	year={2014},
	publisher={Nature Publishing Group}
}

@article{finak2015mast,
	title={{MAST}: {A} flexible statistical framework for assessing transcriptional changes and characterizing heterogeneity in single-cell {RNA} sequencing data},
	author={Finak, Greg and McDavid, Andrew and Yajima, Masanao and Deng, Jingyuan and Gersuk, Vivian and Shalek, Alex K and Slichter, Chloe K and Miller, Hannah W and McElrath, M Juliana and Prlic, Martin and others},
	journal={Genome Biology},
	volume={16},
	number={1},
	pages={278},
	year={2015},
	publisher={BioMed Central}
}

@article{lun2016pooling,
	title={Pooling across cells to normalize single-cell {RNA} sequencing data with many zero counts},
	author={Lun, Aaron TL and Bach, Karsten and Marioni, John C},
	journal={Genome Biology},
	volume={17},
	number={1},
	pages={75},
	year={2016},
	publisher={Springer}
}

@article{li2019bayesian,
	title={Bayesian negative binomial mixture regression models for the analysis of sequence count and methylation data},
	author={Li, Qiwei and Cassese, Alberto and Guindani, Michele and Vannucci, Marina},
	journal={Biometrics},
	volume={75},
	number={1},
	pages={183--192},
	year={2019},
	publisher={Wiley Online Library}
}

@article{zhang2020spatial,
	title={Spatial molecular profiling: {P}latforms, applications and analysis tools},
	author={Zhang, Minzhe and Sheffield, Thomas and Zhan, Xiaowei and Li, Qiwei and Yang, Donghan M and Wang, Yunguan and Wang, Shidan and Xie, Yang and Wang, Tao and Xiao, Guanghua},
	journal={Briefings in Bioinformatics},
	year={2020},
	publisher={Oxford University Press}
}

@article{lubeck2014single,
	title={Single-cell in situ {RNA} profiling by sequential hybridization},
	author={Lubeck, Eric and Coskun, Ahmet F and Zhiyentayev, Timur and Ahmad, Mubhij and Cai, Long},
	journal={Nature Methods},
	volume={11},
	number={4},
	pages={360},
	year={2014},
	publisher={Nature Publishing Group}
}

@article{eng2019transcriptome,
	title={Transcriptome-scale super-resolved imaging in tissues by {RNA} {seqFISH+}},
	author={Eng, Chee-Huat Linus and Lawson, Michael and Zhu, Qian and Dries, Ruben and Koulena, Noushin and Takei, Yodai and Yun, Jina and Cronin, Christopher and Karp, Christoph and Yuan, Guo-Cheng and others},
	journal={Nature},
	volume={568},
	number={7751},
	pages={235--239},
	year={2019},
	publisher={Nature Publishing Group}
}

@article{chen2015spatially,
	title={Spatially resolved, highly multiplexed {RNA} profiling in single cells},
	author={Chen, Kok Hao and Boettiger, Alistair N and Moffitt, Jeffrey R and Wang, Siyuan and Zhuang, Xiaowei},
	journal={Science},
	volume={348},
	number={6233},
	year={2015},
	publisher={American Association for the Advancement of Science}
}

@article{staahl2016visualization,
	title={Visualization and analysis of gene expression in tissue sections by spatial transcriptomics},
	author={St{\aa}hl, Patrik L and Salm{\'e}n, Fredrik and Vickovic, Sanja and Lundmark, Anna and Navarro, Jos{\'e} Fern{\'a}ndez and Magnusson, Jens and Giacomello, Stefania and Asp, Michaela and Westholm, Jakub O and Huss, Mikael and others},
	journal={Science},
	volume={353},
	number={6294},
	pages={78--82},
	year={2016},
	publisher={American Association for the Advancement of Science}
}

@article{rodriques2019slide,
	title={Slide-seq: {A} scalable technology for measuring genome-wide expression at high spatial resolution},
	author={Rodriques, Samuel G and Stickels, Robert R and Goeva, Aleksandrina and Martin, Carly A and Murray, Evan and Vanderburg, Charles R and Welch, Joshua and Chen, Linlin M and Chen, Fei and Macosko, Evan Z},
	journal={Science},
	volume={363},
	number={6434},
	pages={1463--1467},
	year={2019},
	publisher={American Association for the Advancement of Science}
}

@article{svensson2018spatialde,
	title={{SpatialDE}: identification of spatially variable genes},
	author={Svensson, Valentine and Teichmann, Sarah A and Stegle, Oliver},
	journal={Nature Methods},
	volume={15},
	number={5},
	pages={343--346},
	year={2018},
	publisher={Nature Publishing Group}
}

@article{sun2020statistical,
	title={Statistical analysis of spatial expression patterns for spatially resolved transcriptomic studies},
	author={Sun, Shiquan and Zhu, Jiaqiang and Zhou, Xiang},
	journal={Nature Methods},
	volume={17},
	number={2},
	pages={193--200},
	year={2020},
	publisher={Nature Publishing Group}
}

@article{roberts2013gaussian,
	title={Gaussian processes for time-series modelling},
	author={Roberts, Stephen and Osborne, Michael and Ebden, Mark and Reece, Steven and Gibson, Neale and Aigrain, Suzanne},
	journal={Philosophical Transactions of the Royal Society A: Mathematical, Physical and Engineering Sciences},
	volume={371},
	number={1984},
	pages={20110550},
	year={2013},
	publisher={The Royal Society Publishing}
}

@article{diggle1998model,
	title={Model-based geostatistics},
	author={Diggle, Peter J and Tawn, Jonathan A and Moyeed, Rana A},
	journal={Journal of the Royal Statistical Society: Series C (Applied Statistics)},
	volume={47},
	number={3},
	pages={299--350},
	year={1998},
	publisher={Wiley Online Library}
}

@article{edsgard2018identification,
	title={Identification of spatial expression trends in single-cell gene expression data},
	author={Edsg{\"a}rd, Daniel and Johnsson, Per and Sandberg, Rickard},
	journal={Nature Methods},
	volume={15},
	number={5},
	pages={339--342},
	year={2018},
	publisher={Nature Publishing Group}
}

@article{love2014moderated,
	title={Moderated estimation of fold change and dispersion for {RNA-seq} data with {DESeq2}},
	author={Love, Michael I and Huber, Wolfgang and Anders, Simon},
	journal={Genome Biology},
	volume={15},
	number={12},
	pages={550},
	year={2014},
	publisher={Springer}
}

@article{robinson2010edger,
	title={{edgeR}: {A} {Bioconductor} package for differential expression analysis of digital gene expression data},
	author={Robinson, Mark D and McCarthy, Davis J and Smyth, Gordon K},
	journal={Bioinformatics},
	volume={26},
	number={1},
	pages={139--140},
	year={2010},
	publisher={Oxford University Press}
}

@article{moran1950notes,
	title={Notes on continuous stochastic phenomena},
	author={Moran, Patrick AP},
	journal={Biometrika},
	volume={37},
	number={1/2},
	pages={17--23},
	year={1950},
	publisher={JSTOR}
}

@article{li2007beyond,
	title={Beyond {Moran's I}: {T}esting for spatial dependence based on the spatial autoregressive model},
	author={Li, Hongfei and Calder, Catherine A and Cressie, Noel},
	journal={Geographical Analysis},
	volume={39},
	number={4},
	pages={357--375},
	year={2007},
	publisher={Wiley Online Library}
}

@article{stern2000tyrosine,
	title={Tyrosine kinase signalling in breast cancer: {ErbB} family receptor tyrosine kinases},
	author={Stern, David F},
	journal={Breast Cancer Research},
	volume={2},
	number={3},
	pages={1--8},
	year={2000},
	publisher={BioMed Central}
}
\end{document}